\documentclass[aps,prl,longbibliography,reprint]{revtex4-1}%
\usepackage{epsfig,pslatex,latexsym,times,amssymb,amsmath,graphicx}
\usepackage{bm, ulem, float, lipsum}
\usepackage{amsmath}
\usepackage{amsfonts}
\usepackage{amssymb}
\usepackage{mathrsfs}
\usepackage{color}
\usepackage{graphicx}%
\providecommand{\U}[1]{\protect\rule{.1in}{.1in}}
\begin{document}
\author{N. J. Harmon}
\email{nicholas-harmon@uiowa.edu} 
\affiliation{Department of Physics and Astronomy and Optical Science and Technology Center, University of Iowa, Iowa City, Iowa
52242, USA}
\author{M. E. Flatt\'e}
\email{michael\_flatte@mailaps.org} 
\affiliation{Department of Physics and Astronomy and Optical Science and Technology Center, University of Iowa, Iowa City, Iowa
52242, USA}
\date{\today}
\title{Theory of spin-coherent electrical transport through a defect spin state in a metal/insulator/ferromagnet tunnel junction undergoing ferromagnetic resonance}
\begin{abstract}
We describe the coherent dynamics of electrical transport through a localized spin-dependent state, such as is associated with a defect spin, at the interface of a ferromagnet and a non-magnetic material during ferromagnetic resonance.
As the ferromagnet's magnetic moment precesses, charge carriers are dynamically spin-filtered by the localized state, leading to a dynamic spin accumulation on the defect. Local effective magnetic fields modify the precession of a spin on the defect, which also modifies the time-integrated total charge current through the defect. We thus identify a new form of current-detected spin resonance that reveals the local magnetic environment of a carrier spin located at a defect, and thus potentially the defect's identity.
\end{abstract}
\maketitle	

The emerging field of ``quantum spintronics"  seeks to engineer and manipulate single coherent spin systems for the sake of quantum-enhanced sensing/imaging technologies and quantum computing \cite{Awschalom2013}.
Defect spins in an insulating region between a ferromagnetic metal and a nonmagnetic conductor produce an array of coherent spin-dependent phenomena, including defect-associated spin pumping \cite{Wolfe2014,Pu2015,Yue2015},  thermal spin transport \cite{Uchida2010}, and small-field magnetoresistance under electrical bias \cite{Song2014,Txoperena2014,Inoue2015}.  
Individual spin-coherent defects have even been electrically detected in precisely-designed junctions \cite{Baumann2015,Pla2012}.
However, the potential of a coherently-precessing source of spins, readily available from a ferromagnetic contact undergoing precession (such as from a spin torque oscillator) has not yet been explored; such a coherent source may be able to reach a {\it single-defect-spin regime} of spin pumping or dynamic spin polarization.

Here we predict observable coherent dynamics in the charge and spin transport through a single defect in the junction between a ferromagnetic material and a second, nonmagnetic (NM) conducting material, when the magnetism of the ferromagnet (FM) precesses in time such as during 
 ferromagnetic resonance (FMR). 
 During electrical transport the defect  can become dynamically spin polarized, and its  spin manipulated, even with negligible  coupling between the defect and FM from a magnetic dipolar field or exchange interaction. This provides a single-defect-spin example of dynamic spin polarization.
Analysis of the current through the device reveals the local spin character of a defect and its environment  without the need of a microwave cavity. These effects, in the single-defect limit, would be detectable with a spin-polarized scanning tunneling microscope tip undergoing FMR, and should persist even for sequential hopping transport between the tip and the defect, as well as between the defect and the second conducting contact. A slower transport rate between the defect and the FM  provides better resolution of the defect's local environment, so long as the defect spin state's coherence time is comparable to or exceeds the electrical transport rate through the junction.

\begin{figure}[ptbh]
 \begin{centering}
        \includegraphics[scale = 0.275,trim = 0 0 0 0, angle = -0,clip]{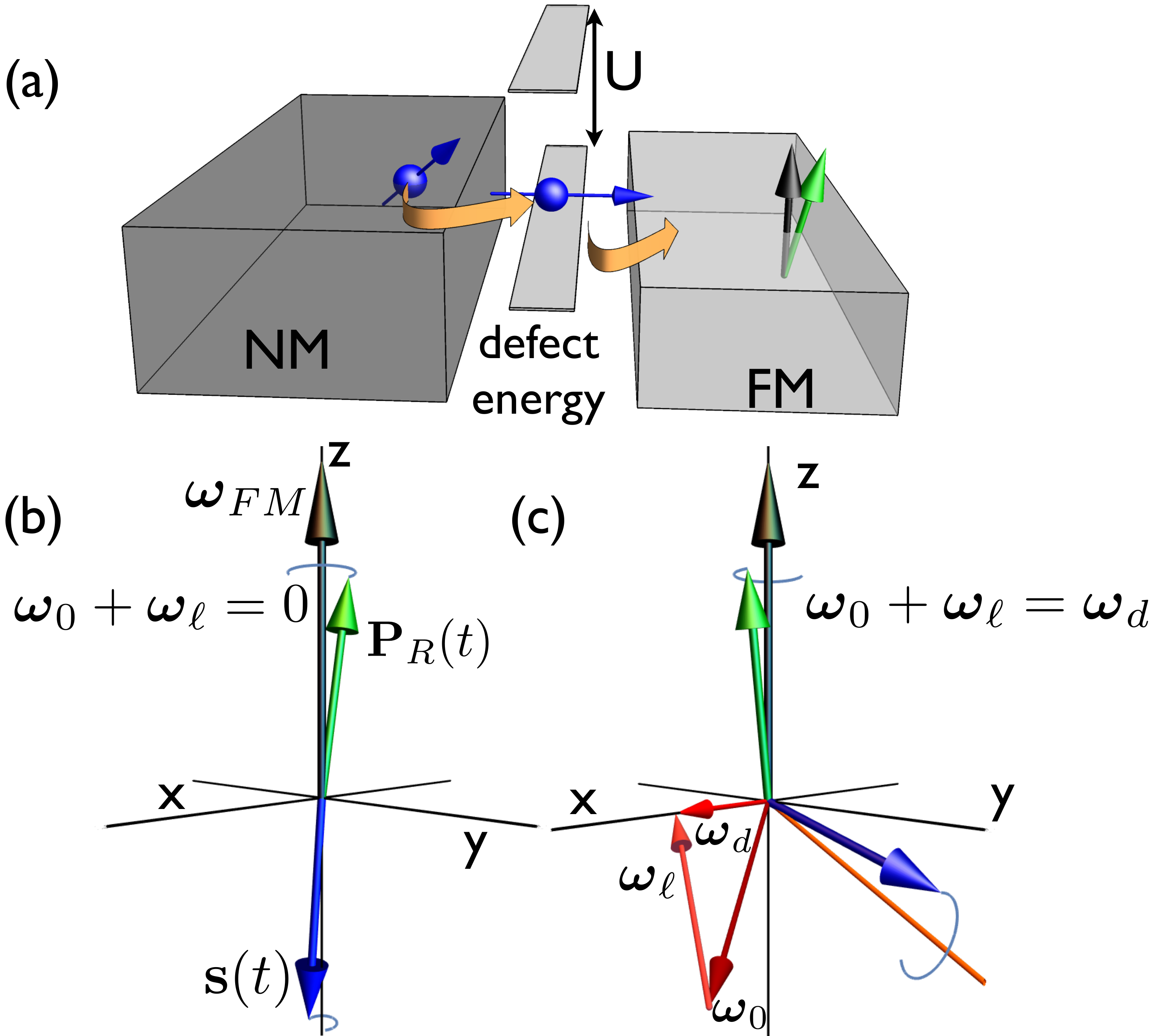}
        \caption[]
{(a) Diagram of the energy landscape of a ferromagnet/nonmagnetic  (FM/NM) metal junction. The darker box specifies the NM metal and the middle planes represent the two energy levels of the defect which are separated by an on-site Coulomb energy $U$. The bias pushes electrons through the junction  from the NM metal (left) to the FM (right). The vertical direction is energy whereas the lateral directions are spatial coordinates. (b) Schematic of the spatial orientation of various spins: the FM's polarization, $\bold{P}_R(t)$ (green arrow) precesses around an axis $\boldsymbol{\omega}_{FM}$  (black arrow). The spin of the defect $\bm{s}(t)$ (blue arrow) precesses in the sum of an externally applied and a local magnetic field, at a frequency $\bm{\omega}_d = \bm{\omega}_0 + \bm{\omega}_{\ell}$. In this panel $\bm{\omega}_d = 0$. The dynamical spin polarization of the defect follows the FM's polarization. (c) For $\bm{\omega}_d \ne 0$ the defect spin precesses around the static ($\omega_{FM} =0$) steady state spin orientation (indicated by the orange line).}\label{fig:impurityTypes}
        \end{centering}
\end{figure}

Here we focus on a defect electronic structure corresponding to a single orbital state and two (oppositely-oriented) spin states, either unoccupied or singly occupied by a spin-$1/2$ electron.  
The junction is shown schematically in Fig.~\ref{fig:impurityTypes}. 
Transport occurs as an electron spin, of arbitrary direction,  hops from the left contact to the previously empty defect site and singly occupies the level. The electron's subsequent motion will then be limited  depending on the  orientation of its spin relative to the majority spin polarization at the Fermi level in the FM; if parallel then the transport is rapid, while if antiparallel the transport is slower. Similar behavior will occur for hole spin transport, with opposite bias voltage and when the hole hops to a defect site that is empty (of holes, and thus doubly-occupied by electrons), or for defects with different electronic state ordering, so long as the transport through the defect states  depends on spin.  For example, a ground-state spin-$1$ defect, such as a silicon carbide divacancy \cite{Koehl2011}, will exhibit essentially the same features as our spin-1/2 system, but with opposite dynamic spin polarization. We focus on the case shown in Fig.~\ref{fig:impurityTypes}. 

A heuristic picture helps visualize the resonance condition for transport through the defect state during precession of the FM's spin polarization.
The spin polarization of the FM's Fermi-level carriers, $\bm{P}_R(t)$ (green arrow), precesses around an axis  $\bm{\omega}_{FM}$ (black arrow), depicted parallel to $\hat{z}$ in Fig.~\ref{fig:impurityTypes}.  The cone angle is the angle  between $\bm{P}_R(t)$ and $\bm{\omega}_{FM}$.
The equilibrium polarization of the FM when not undergoing FMR is $\bm{P}_R || \hat{z}$.
The probability for a carrier at the defect to enter the FM depends on the relative orientation of the carrier's spin, $\bm{s}(t)$ (blue arrow), and  $\bm{P}_R(t)$. For the simplest picture consider the FM to be 100\%\ spin polarized, for which only a carrier with some spin component parallel to $\bm{P}_R(t)$ may tunnel into the FM. 

The spin on the defect site, associated with the carrier, can also precess due to the influence of an applied magnetic field as well as a local effective field arising from hyperfine interactions, exchange interactions with neighboring sites, or other effects. The directions of precession vectors will be described using a polar angle $\theta$ relative to $\bm{\omega}_{FM} \parallel \hat z$ and an azimuthal angle $\phi$ relative to the $\hat x$ axis, with a subscript corresponding to the specific precession vector.
The local field is considered to be independent of the applied magnetic field, and causes the defect spin to precess according to the precession vector $\bm{\omega}_\ell$. The applied magnetic field precesses the defect spin according to the precession vector $\bm{\omega}_0$, and the total precession will be  $\bm{\omega}_d = \bm{\omega}_0 + \bm{\omega}_{\ell}$. To distinguish this precession frequency from apparent precession due to spin filtering, the precession frequency $\omega_d$ will be referred to as the {\it defect spin's Larmor frequency}.

\begin{figure}[ptbh]
 \begin{centering}
        \includegraphics[scale = 0.5,trim = 0 0 0 0, angle = -0,clip]{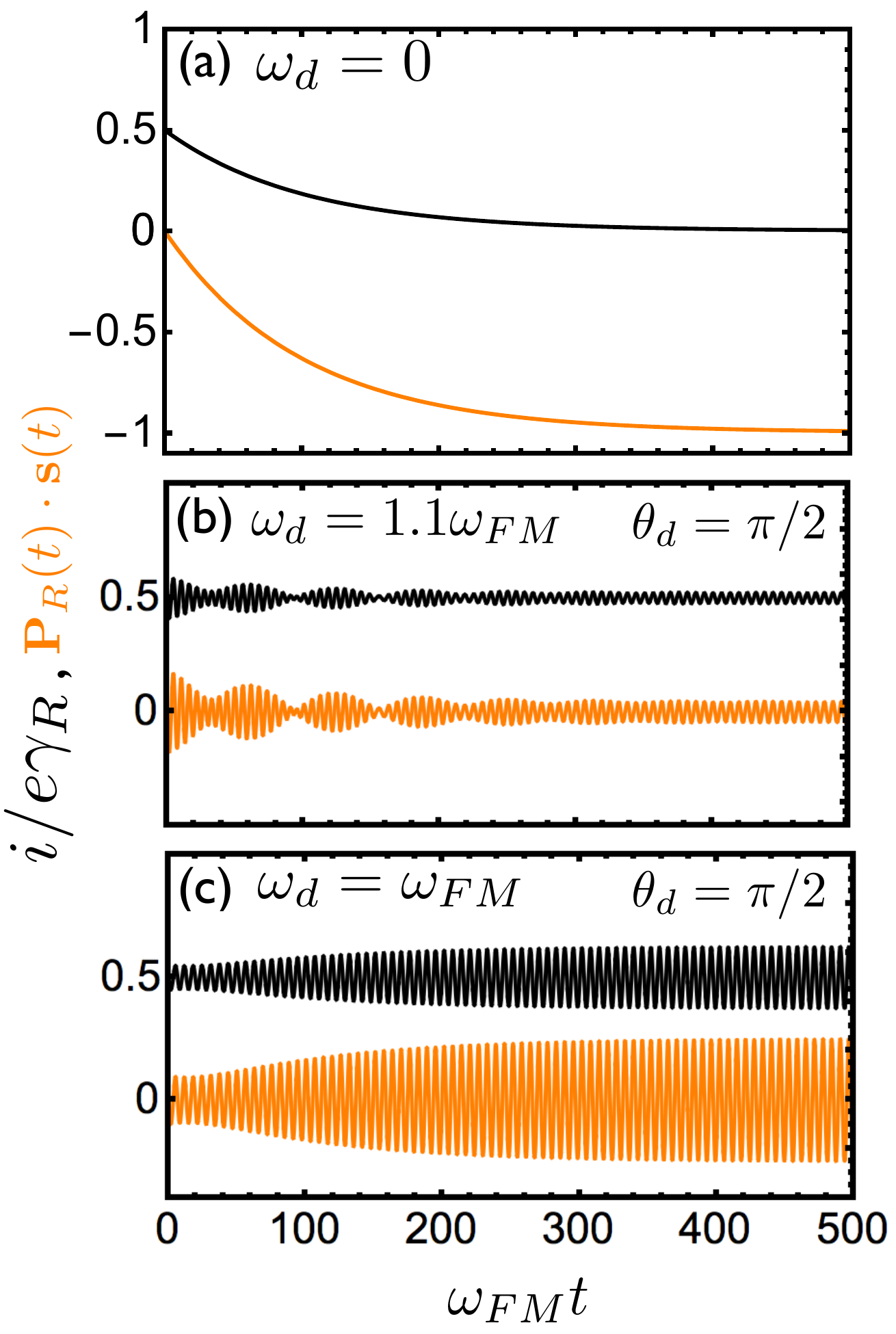}
        \caption[]
{(a) Charge current [Eq.~(\ref{eq:current})]
when the defect spin's  precession frequency, $\omega_d$, is zero. 
The current (black line) decreases to zero as the carrier spin at the defect (orange line) becomes polarized opposite that of the FM. Once the defect is completely antiparallel, no further charge can occupy or leave the defect.
(bc) Charge current from two choices of  $\omega_d$, (b) non-resonant and (c) resonant, with $\bm{\omega}_d$ oriented along the $x$-axis in Fig.~\ref{fig:impurityTypes}. %($\theta = \pi/2$). 
The orange curves depict the projection of the carrier spin, $\bold{s}(t)$, onto the rotating polarization, $\bold{P}_R(t)$, which determines the current (black lines). Parameters are $\phi_d = 0$, cone angle between $\bm{P}_R(t)$ and $\bm{\omega}_{FM}$ of $0.05$~radians ($\sim 10\%$),  $\gamma_L = 10{\omega}_{FM}$ and $\gamma_R = 0.01{\omega}_{FM}$, $P_R = 1$, and $P_L = 0$. 
For clarity, each amplitude is enhanced by a factor of 10.
}\label{fig:currentVsTime} 
        \end{centering}
\end{figure}

{\it Dynamic spin polarization} emerges on the defect site, and is largest when $\bm{\omega}_d=0$,  shown in Fig.~\ref{fig:impurityTypes}(b).  Under bias the defect occupation is continually replenished, until the carrier spin on the defect is oriented {\it antiparallel} to $\bm{P}_R$ and no further transport occurs until the carrier spin decoheres or the FM polarization changes. This spin filtering process results in the defect spin tracking approximately antiparallel to $\bm{P}_R(t)$,  therefore blocking the current through the junction. Figure~\ref{fig:currentVsTime} illustrates the details of the spin-coherent effects on charge current during FMR, beginning with an unoccupied defect spin state. Figure~\ref{fig:currentVsTime}(a) demonstrates (orange line) that $\bm{s}(t) \cdot \bm{P}_R(t) \rightarrow - 1$ after  transient dynamics.

Figure~\ref{fig:impurityTypes}(c) shows the changing dynamics for a non-vanishing Larmor precession of the carrier spin on the defect, and for $\bm{\omega}_d$ perpendicular  to $\bm{\omega}_{FM}$. For $\bm{\omega}_{FM}=0$ the defect spin precession causes the dynamic spin polarization generated from spin filtering in transport into the FM to rotate in the $zy$ plane and be oriented along the orange line, which is determined by the relative precession frequency and spin filtering rate to be
 \begin{equation}
\bm{s}(t) = - \frac{2\gamma_L[\gamma_R^2 \bm{P}_R + \gamma_R \bm{\omega}_d\times \bm{P}_R + (\bm{\omega}_d \cdot \bm{P}_R) \bm{\omega}_d]}{(\gamma_R (1 - P_R^2 \chi(\bm{\omega}_d)) + 2 \gamma_L)(\gamma_R^2 + \omega_d^2)},
 \end{equation}
with
 \begin{equation}
\chi(\bm{\omega}_d) = \frac{\gamma_R^2  + (\bm{\omega}_d\cdot \hat{P}_R)^2}{\gamma_R^2 + \omega_d^2},
 \end{equation}
where  $\gamma_L$ is the  hopping rate from the left conductor to the defect and $\gamma_{R}$ the hopping rate from the defect to the FM. 

For $\bm{\omega}_{FM}\ne 0$ the 
dynamical defect spin polarization $\bm{s}(t) $ precesses at the frequency ${\omega}_{FM}$ around the orange line, as indicated in Fig.~\ref{fig:impurityTypes}(c).  
Figure~\ref{fig:currentVsTime} displays the current for this configuration off resonance [Fig.~\ref{fig:currentVsTime}(b)] and on resonance [Fig.~\ref{fig:currentVsTime}(c)]. When off resonance  some beating occurs in the transient stage until the defect spin $\bm{s}(t)$ is syncronized with $\bm{P}_R(t)$ \footnote{A brief animation of the magnetization and defect spin dynamics is found in the online Supplementary Information.}. 
On resonance, corresponding to $\omega_{FM} = |\bm{\omega}_{0} + \bm{\omega}_{\ell}| =  \omega_d$, the amplitudes of the defect spin's 
precession and the current oscillations increase. 

 \begin{figure}[ptbh]
 \begin{centering}
        \includegraphics[scale = 0.5,trim = 0 0 0 0, angle = -0,clip]{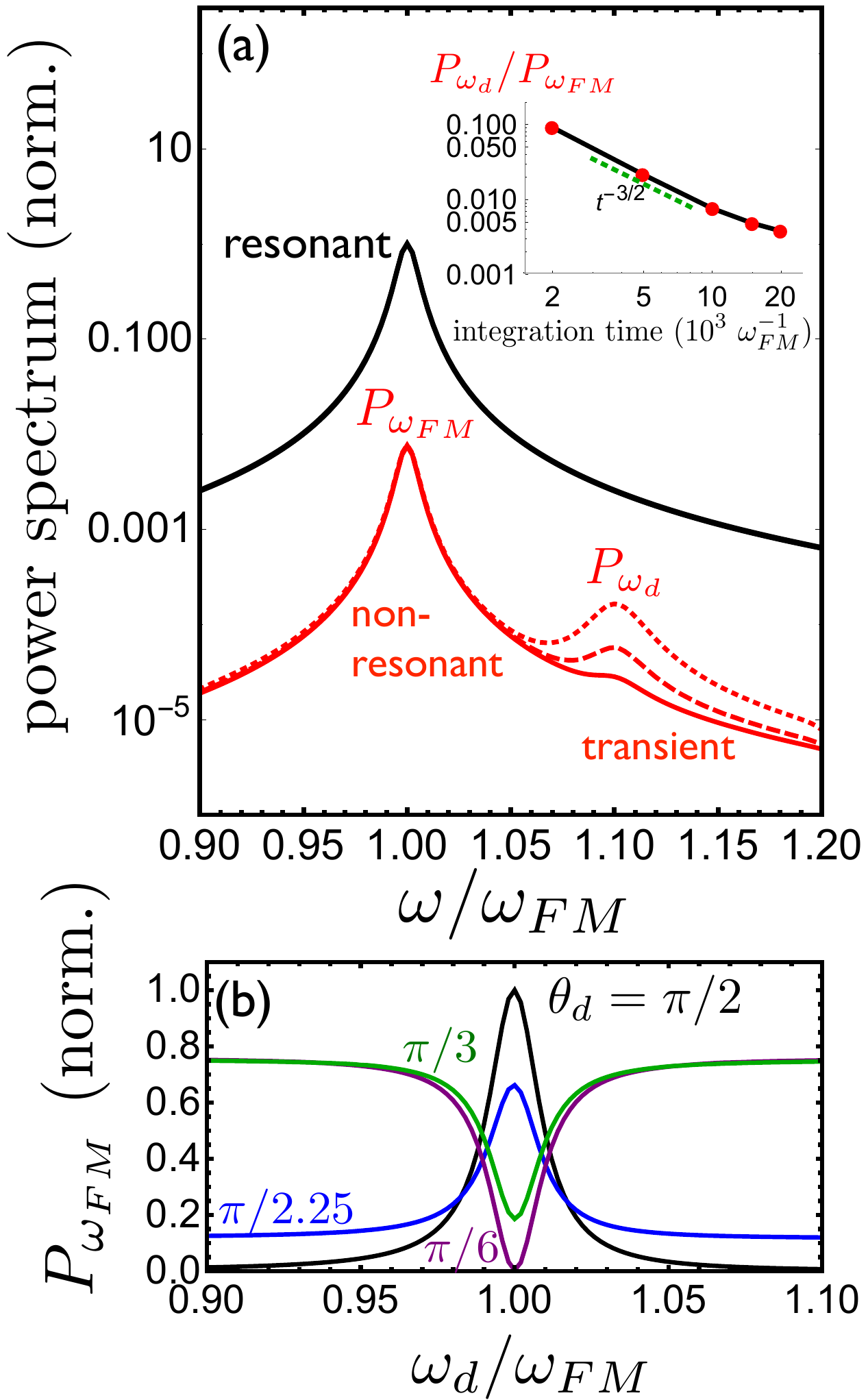}
        \caption[]
{
(a) Power spectra for the resonant (black) and non-resonant (red) currents in Fig.~\ref{fig:currentVsTime}, each normalized with respect to the resonant peak. The off-resonant spectrum shows a transient peak at
$\omega_d=1.1\omega_{FM}$, in addition to a persistent peak at $\omega_d=\omega_{FM}$ with width governed by a damping rate $\Gamma = 0.005\omega_{FM}$. Integration times are 5 (dotted), 10 (dashed), and 20 (solid)  $\times 10^3$ $\omega_{FM}^{-1}$. 
(Inset) dependence on the integration time of the off-resonant ratio of the power at the Larmor frequency to that at the FMR frequency  $(P_{\omega_{d}}/P_{\omega_{FM}})$.
(b) $P_{\omega_{FM}}$  versus $\omega_d$, for several  angles ($\theta_d$) between $\bm{\omega}_d$ and $\bm{\omega}_{FM}$. $P_{\omega_{FM}}$ is  independent of $\phi_d$. Parameters are identical to those  in Fig. \ref{fig:currentVsTime}.}\label{fig:powerspec} 
        \end{centering}
\end{figure}

The  off-resonant power spectrum (red) of the current oscillations, Fig.~\ref{fig:powerspec}(a), shows peaks  at both the FMR frequency ($\omega_{FM}$) and the defect spin's precession frequency ($\omega_d$); the peak at $\omega_d$ is a transient, as shown with  integration times of 5, 10, 20  $\times 10^3$ $\omega_{FM}^{-1}$, and in the inset.
When on resonance (black), $\omega_{FM} = |\bm{\omega}_{0} + \bm{\omega}_{\ell}| =  \omega_d$,
$\bold{s}(t)$ and $\bold{P}_R(t)$
 are synchronized and $s(t)$ increases, producing larger amplitude current oscillations. 
Figure~\ref{fig:powerspec}(b) shows the dependence of the current's power spectrum at the FMR frequency, $P_{\omega_{FM}}$,  on $\omega_d$ for several different  orientations $\theta_d$.

We now describe how the charge current through the junction during FMR is calculated including the spin-coherent dynamics of the defect. The current operators involving the two contacts, from the NM contact to the defect (`left' current), and from the defect to the FM (`right' current), are explicitly constructed and combined with a coherent density matrix treatment of the carrier spin dynamics. 
The following ansatz describes the `right' current operator
\begin{equation}\label{eq:iR}
 \hat{i}_R(t) = \frac{e}{2} \gamma_R \Big[\hat{\bold{\mathcal{P}}}_{R}(t) \rho(t) + \rho^{\dagger}(t) \hat{\bold{\mathcal{P}}}^{\dagger}_{R}(t) \Big],
 \end{equation}
where $\mathcal{P}_{R}(t)$ is the polarization operator of the FM and $\rho(t)$ the density matrix of the defect's carrier spin. The second term of Eq.~(\ref{eq:iR}) ensures hermiticity. $\hat{\bold{\mathcal{P}}}_{R}(t) = \frac{1}{2}(I + \bold{P}_R(t)\cdot \bold{\sigma})$ describes an imperfect spin filter ($\hat{\bold{\mathcal{P}}}_{R}(t)$ is not idempotent unless $P_R= 1$) \cite{Farago1971}.
$\bm{P}_R$, determined by $\text{Tr}(\hat{\bold{\mathcal{P}}} \bold{\sigma})$, precesses around $\bm{\omega}_{FM}$ and is determined by 
\begin{equation}
\dot{\hat{\bold{\mathcal{P}}}}_{R}(t) = -\frac{1}{2}\frac{i}{\hbar}[\hbar \bm{\omega}_{FM} \cdot \bm{\sigma}, \hat{\bold{\mathcal{P}}}_{R}(t)]. 
\end{equation}
An analytic solution for $\hat{\bold{\mathcal{P}}}_{R}(t)$ is available using an algebraic solver \footnote{The explicit form of the matrix $\hat{\bold{\mathcal{P}}}_{R}(t)$  is found in the Supplementary Information}.  
To account for the finite line width of the FMR, the power spectrum is convolved with a Lorentzian function of width $\Gamma$ \footnote{A description of the convolution is found in the Supplementary Information}. 

$\hat{i}_R$ represents the movement of charge combined with spin information encoded in the matrix elements.
Charge (spin) current is $i_R =\text{Tr}\hat{i}_R$ ($\bm{i}_{s, R}=\text{Tr}\hat{i}_R\bm{\sigma}$). 
The right charge current once the defect site is filled,
\begin{equation}\label{eq:current}
i = \text{Tr}(\hat{i}_R)= \frac{1}{2} ( 1 + \bm{s}(t) \cdot \bm{P}_R(t)) e \gamma_R,\quad \text{with } \bm{s} =  \text{Tr}(\rho \boldsymbol{\sigma}),
\end{equation}
 which illustrates the dependence of the current on the relative alignment of the defect spin and FM's polarization. For  $\gamma_L \gg \gamma_R$  the defect state is predominately filled. For a spin-polarized contact that is an STM tip, the tip can be moved away from the impurity until $\gamma_L \gg \gamma_R$.
 The amplitude of  current oscillations, for small cone angles and $\gamma_L \gg \gamma_R$, scales as $P_R^2$. 
 
 The `left' current (NM contact to defect) can be derived in a similar fashion after constraining the defect to be at most singly occupied. For a left conductor with a static magnetization,
 \begin{equation}\label{eq:iL}
\hat{i}_L(t) = e\gamma_L[1-\text{Tr}\rho(t)]\hat{\bold{\mathcal{P}}}_{L},
\end{equation}
where $\hat{\bold{\mathcal{P}}}_{L}= \frac{1}{2}(I + \bold{P}_L\cdot \bold{\sigma})$ is the polarization operator of the left conductor. This formalism can be generalized to include dynamic magnetization of the left conductor, although here we present results
only for a NM, {\it i.e.}  ${\bold{P}}_{L}= 0$.
Charge conservation demands that the `left' current be the same as the `right' current for a time-independent $\bm{P}_R(t)$ or when the current is averaged over a precession period of $\bm{P}_R(t)$, so $\overline{\text{Tr}\hat{i}_L} = \overline{\text{Tr}\hat{i}_R}$. 

Construction of the defect density matrix $\rho(t)$ consistently connects  the    currents and determines their sensitivity to spin and applied magnetic fields. 
The stochastic Liouville equation is suited well for this type of problem \cite{Kubo1963, Haberkorn1976}, so 
\begin{equation}\label{eq:SLE}
\dot{\rho}(t) = -\frac{i}{\hbar} [\mathscr{H} , \rho(t)] -  \gamma_R \{ \hat{\bold{\mathcal{P}}}_{R}(t), \rho(t)\} 
 +  2\gamma_L [1 - \text{Tr}\rho(t)] %\hat{\bold{\mathcal{P}}}_{L}
 .
 \end{equation} 
The first term of Eq.~(\ref{eq:SLE}) produces the coherent evolution of the spin, the second term (curly braces are anti-commutators)  the spin-selective nature of tunneling into the FM. The last term describes hopping onto the defect site from the left contact. 
 $\mathscr{H}= (\hbar/2)\bm{\omega}_d \cdot \bm{\sigma}$ is the spin Hamiltonian at the defect site.  In typical insulators the localization length of the defect's wave function is wide enough to encompass a large number of randomly oriented nuclei, so  a  local hyperfine field $\bm{\omega}_\ell$ can be accurately approximated as a classical vector.  The spin density matrix is obtained from a numerical solution to  Eq.~(\ref{eq:SLE}), and the  current from either Eq.~(\ref{eq:iR}) or Eq.~(\ref{eq:iL}).

\begin{figure}[ptbh]
 \begin{centering}
        \includegraphics[scale = 0.375,trim = 0 0 0 0, angle = -0,clip]{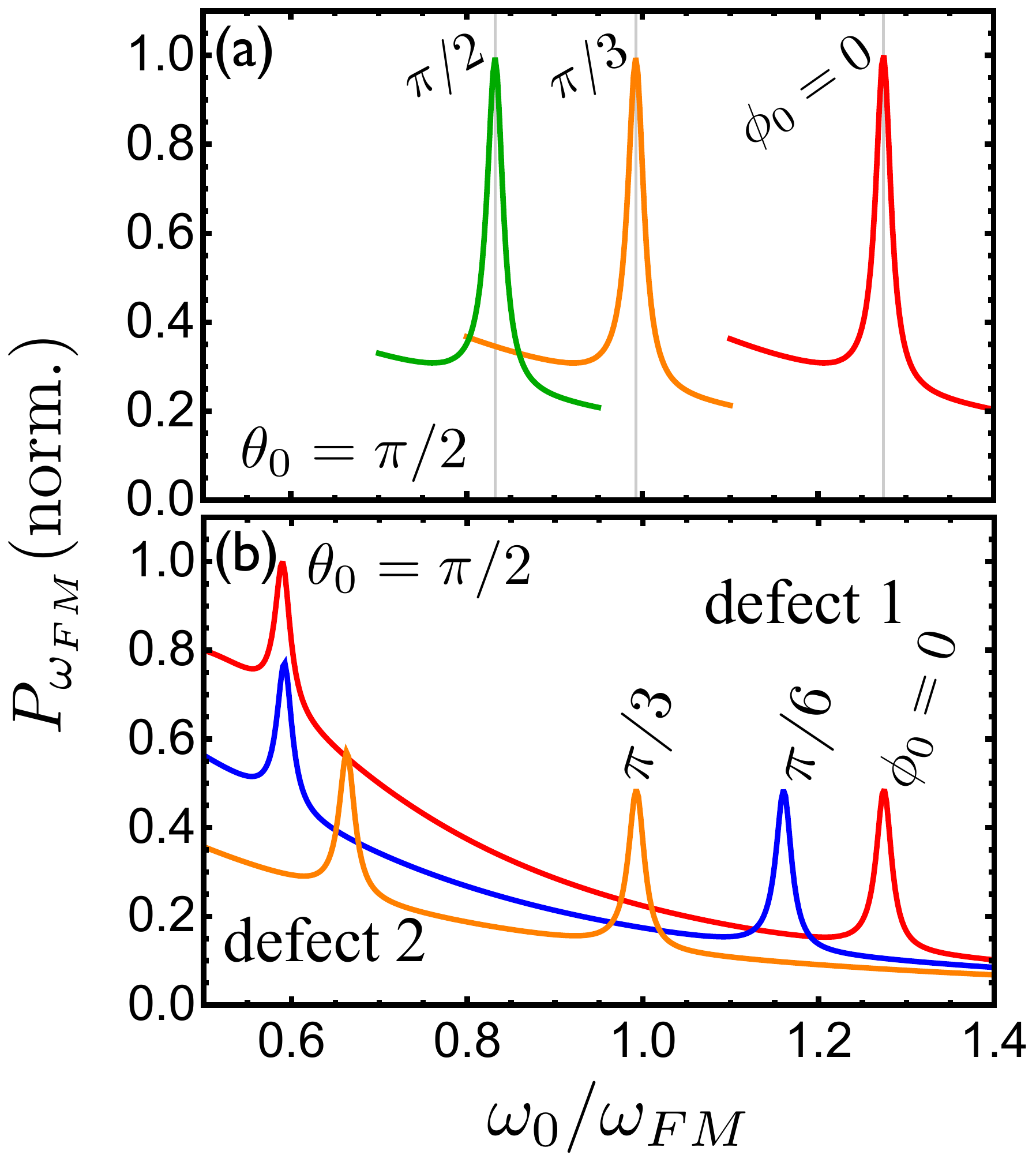}
        \caption[]
{(a) Plots showing the integrated current at $\omega_{FM}$ when the applied field, $\omega_0$, is swept.  Resonances occur when $\omega_d = |\bm{\omega}_0 + \bm{\omega}_{\ell}| = \omega_{FM}$. Here $\bm{\omega}_{\ell} = (-0.3,  0.1, 0.2)\omega_{FM}$. 
(b) Two resonance features appear when two defects are probed.  Each colored curve corresponds to an independent sweep of the magnetic field in the $x-y$ plane at an angle $\phi_0$.
For the two defects, $\bm{\omega}_{\ell,1} = (-0.3, 0.1, 0.2)\omega_{FM}$ and $\bm{\omega}_{\ell,2} = (0.4, 0.1, -0.1)\omega_{FM}$.
Curves in (a) and (b) are normalized to the highest peak and labeled by the applied field's azimuthal angle $\phi_0$.
Parameters are identical to those used in Fig. \ref{fig:powerspec}.
}\label{fig:1defect} 
        \end{centering}
\end{figure}

Although the resonances always occur  when $\omega_{FM} = \omega_d$, independent of precession axis direction, it is possible to determine $\bm{\omega}_\ell$ by measuring  ${\omega}_0$ at resonance, for several different directions of $\bm{\omega}_0$, as $\omega_0$ at resonance will vary with direction from $ \omega_{FM}-\omega_\ell$ to $\omega_{FM}+\omega_\ell$.
In Fig.~\ref{fig:1defect}(a) the current's power spectrum at the FMR frequency, $P_{\omega_{FM}}$, is shown as a function of $\omega_0$ for three different directions of $\bm{\omega}_0$, for an example hyperfine local field  $\bm{\omega}_{\ell} = (-0.3, 0.1, 0.2)\omega_{FM}$. This theory applies also to two independent defects through which parallel currents run. Fig.~\ref{fig:1defect}(b) displays  sweeps of $\omega_0$ at $\theta_0=\pi/2$ and three different $\phi_0$, similar to the single defect scenario. 
Now resonances occur at two different applied fields for each sweep \footnote{Further description of the resonant detection is located in the Supplementary Information.}.

$\omega_{FM}$ is fixed in Figs.~\ref{fig:powerspec}(b) and~\ref{fig:1defect} as $\omega_0$ varies. For a ferromagnetic  thin film with the easy axis of the contact in the film plane, the component of the applied magnetic field along the hard axis, if sufficiently small, does not influence  $\bm{\omega}_{FM}$ but does change $\bm{\omega}_0$ and $\bm{\omega}_d$. We assume the magnetic field component along the hard axis is varied in order to vary $\bm{\omega}_d$ leaving $\bm{\omega}_{FM}$ fixed.

The relevant timescale for differential precession of the carrier spin and the FM is the timescale for hopping from the defect to the FM.
For typical scanning tunneling microscopy measurements with currents of $0.1-30$~nA \cite{Chenbook}, the timescale for hopping from a defect to a ferromagnetic tip would be 0.05--1.6~ns. 
For spins on the defect coherent on this timescale, which is known to be the case for many examples of localized spins \cite{Koenraad2011}, the features described here will emerge. 
By comparing this hopping time to the precession time of the carrier spin on the defect in a local magnetic field, the sensitivity to local fields can be estimated to be of the order of $\sim 10$~mT, characteristic of hyperfine fields for many types of defects. Smaller currents will improve sensitivity to $\bm{\omega}_\ell$.

Spin-coherent evolution of a carrier spin at a defect produces resonant features in the charge conductivity of a ferromagnet/insulator/nonmagnet junction.  From this, small numbers of defects, or a single defect, can be identified by matches between the ferromagnetic resonance frequency of a contact and the local  precession of the spin(s) of the defect(s). The approaches described here would also permit the preparation of specific desired defect spin states through appropriate choices for the ferromagnet's precession frequency, leading to controlled studies of the coupled dynamics of two coherent spins.

\begin{acknowledgments}
This work was supported  by the U.S. Department of Energy, Office of Science, Office of Basic Energy Sciences, under Award \#DE-SC0016447.
\end{acknowledgments}

%\bibliography{central-bibliography}
%\bibliography{../../central-bibliography}


\end{document}